\documentclass[twocolumn,prb,amsmath,amssymb,amsfonts,superscriptaddress,floatfix,aps,showpacs,notitlepage]{revtex4-1}

\usepackage{graphicx}
\usepackage{epstopdf}
\usepackage{dcolumn}
\usepackage{bm}
\usepackage{rotating} 
\usepackage{color}
\usepackage{subfigure}
\usepackage{natbib}
\usepackage{enumitem}
\usepackage[Symbol]{upgreek}
\usepackage[symbol*]{footmisc}
\usepackage{lipsum}
\usepackage{amsmath}                    
\usepackage{amssymb}                    

\usepackage{textcomp}					

\begin{document}
\newcommand{\oscar}[1]{\textcolor{red} {#1}}
\renewcommand{\thetable}{\arabic{table}}

\title{Nonconventional screening of the Coulomb interaction in 
Fe$_{x}$O$_{y}$ clusters: \\ An \emph{ab-initio} study}
\date{\today}

\author{L. Peters}
\email{L.Peters@science.ru.nl}
\affiliation{Institute for Molecules and Materials, Radboud University 
Nijmegen, NL-6525 AJ Nijmegen, The Netherlands}

\author{E. \c{S}a\c{s}{\i}o\u{g}lu}
\affiliation{Institut f\"ur Physik, Martin-Luther-Universit\"at Halle-Wittenberg, D-06099 Halle (Saale) Germany}
\affiliation{Peter Gr\"unberg Institut and Institute for Advanced Simulation, Forschungszentrum J\"ulich and JARA, 
52425 J\"ulich, Germany}

\author{S. Rossen}
\affiliation{Institute for Molecules and Materials, Radboud University Nijmegen, NL-6525 AJ Nijmegen, The Netherlands}
\affiliation{Peter Gr\"unberg Institut and Institute for Advanced Simulation, Forschungszentrum J\"ulich and JARA, 
52425 J\"ulich, Germany}

\author{C. Friedrich}
\affiliation{Peter Gr\"unberg Institut and Institute for Advanced Simulation, Forschungszentrum J\"ulich and 
JARA, 52425 J\"ulich, Germany}

\author{S. Bl\"ugel}
\affiliation{Peter Gr\"unberg Institut and Institute for Advanced Simulation, Forschungszentrum J\"ulich and 
JARA, 52425 J\"ulich, Germany}

\author{M. I. Katsnelson}
\affiliation{Institute for Molecules and Materials, Radboud University Nijmegen, NL-6525 AJ Nijmegen, 
The Netherlands}

\begin{abstract}

From microscopic point-dipole model calculations of the screening of the Coulomb interaction 
in non-polar systems by polarizable atoms, it is known that screening strongly depends on 
dimensionality. For example, in one dimensional systems the short range interaction is screened, 
while the long range interaction is anti-screened. This anti-screening is also observed in some 
zero dimensional structures, i.e. molecular systems. By means of ab-initio calculations in 
conjunction with the random-phase approximation (RPA) within the FLAPW method we study screening 
of the Coulomb interaction in Fe$_{x}$O$_{y}$ clusters. For completeness these results are compared with their bulk counterpart magnetite. It appears that the onsite Coulomb interaction is very well screened both in the clusters and bulk. On the other hand for the intersite Coulomb interaction the important observation is made that it is almost contant throughout the clusters, while for the bulk it is almost completely screened. More precisely and interestingly, in the clusters anti-screening is observed by means of ab-initio calculations.   

\end{abstract}

\maketitle


\section{Introduction}

The huge interest in nano-technology is fuelling the trend of downscaling devices. Naturally 
this will reach the regime of small clusters. However, also from a fundamental point of view clusters 
are very interesting. In general clusters behave completely different from their bulk counterpart. In particular, 
the removal or addition of just one atom can change the electronic and magnetic properties 
completely.\cite{jena,chris,lars1,lars2} This clearly provides a huge playground for the design of new devices.

For an efficient design, a proper fundamental understanding of the system is essential. This is 
usually complicated by correlation effects that inhibit an exact solution to the problem. Therefore, 
in practice approximate methods have to be considered. In order to find a proper method, knowledge 
of the correlation effects is crucial. For example, for weakly correlated systems it is known that density 
functional theory (DFT) works very well, while for strong local correlations a (generalized) Hubbard 
model provides a good description. Actually it is the gradient of the (screened) Coulomb interaction that 
matters.\cite{mot1} For a very small gradient, i.e. an almost constant effective Coulomb interaction, 
clearly a mean-field approach and thus single-particle approach is justified. On the other hand, for a very large gradient, i.e., for 
only a local effective Coulomb interaction, the Hubbard model becomes adequate.

It is this important information on the effective Coulomb interaction that is provided in this work for 
the Fe$_{x}$O$_{y}$ clusters. More precisely,  Fe$_{2}$O$_{3}$, Fe$_{3}$O$_{4}$ 
and Fe$_{4}$O$_{6}$ are selected since they are well studied in literature.~\cite{std1,std2,std3,std4,std5,std6,std7,std8,std9} Furthermore, two of them are anti-ferromagnetic, while the other is ferromagnetic. There exist several methods to calculate the effective Coulomb interaction. 
For example, in the bulk usually a uniform dielectric theory can be used.\cite{modb1} Here the system is 
modeled as a continuum and the (\textbf{q}-dependent) dielectric constant is obtained within a mean-field 
approximation. For example, the Clausius Mossotti approximation can be used for ionic insulators. Since 
the dielectric constant depends only on the distance between the charges (and not the crystal structure), 
this approximation is only good when local field corrections can be neglected. However, it is well known 
that these local field corrections become important for low dimensional systems. The microscopic point-dipole 
model can be used to take local field corrections into account.\cite{mot1,modb1} This method is based on the 
assumption that the charge distribution of a polarized system can be considered as a collection of localized 
point dipoles. This assumption works reasonably for localized charge distributions like in ionic insulators, 
but becomes inadequate for systems with delocalized charges due to for example covalent bonds. Since it is 
not clear from the beginning to which regime  Fe$_{x}$O$_{y}$ clusters belong, we use \textit{ab-initio} 
theory in conjunction with the random phase approximation (RPA). In this way also local field corrections 
are included.

Iron oxide clusters and nanoparticles have applications in catalysis, magnetic data storage and biomedical 
treatment due to their unique catalytic, magnetic and biochemical properties.\cite{ap1,ap2,ap3} Furthermore, 
iron oxide interactions are interesting in general for corrosion and biological oxygen transport processes. 
Thus, a detailed understanding of Fe$_{x}$O$_{y}$ clusters could contribute to a better understanding of such 
processes and new technological applications. Due to this interest there have been a number of experimental 
and theoretical studies.~\cite{std1,std2,std3,std4,std5,std6,std7,std8,std9} Most theoretical studies are 
performed with DFT and focus on the geometric structure. From a comparison of the experimental and calculated 
vibrational spectrum the structure of some Fe$_{x}$O$_{y}$ clusters is well established.\cite{std7,std8} 
Furthermore, some studies address in some detail the electronic and magnetic structure. However, to our 
knowledge a detailed consideration of correlation and screening effects does not exist. In our opinion such an understanding
is crucial and should form the basis in determining which methods to use for further studies.

The aim of the present work is the ab-initio determination of the screened Coulomb interaction in Fe$_{2}$O$_{3}$, Fe$_{3}$O$_{4}$ and Fe$_{4}$O$_{6}$ clusters. Employing the random-phase approximation (RPA)  
within the full-potential linearized augmented plane wave (FLAPW) method using Wannier functions
we show that in these clusters the onsite Coulomb interaction is well screened, while the intersite 
Coulomb interactions are barely screened or even anti-screened. The important consequence being that the Coulomb interaction is almost constant throughout the clusters. For completeness we compared these results with their bulk 
counterpart magnetite. Herein only the onsite Coulomb interaction is appreciable, while the intersite 
Coulomb interactions are almost completely screened. The rest of the paper is organized as follows. 
The method and computational details are presented in Section\,\ref{section-2} and Section\,\ref{section-3}, 
respectively. Section\,\ref{section-4} deals with the results and discussion and finally in Section\,\ref{section-5}  
we give the conclusions.

\section{Method} 
\label{section-2}
In this work we study partially and fully screened Coulomb interaction parameters calculated with the ab-initio 
cRPA and RPA methods, respectively. The non-interacting reference system is taken from a preceding DFT calculation.

The effective Coulomb interaction is defined as 
\begin{equation}
W(\boldsymbol{r},\boldsymbol{r}',\omega)=\int d\boldsymbol{r}''  \epsilon^{-1}(\boldsymbol{r},\boldsymbol{r}'',\omega) v(\boldsymbol{r}'',\boldsymbol{r}'),
\label{fullysw}
\end{equation}
where $\epsilon(\boldsymbol{r},\boldsymbol{r}'',\omega)$ is the dielectric function and $v(\boldsymbol{r}'',\boldsymbol{r}')$ is the bare Coulomb interaction potential. Since an exact expression for the dielectric function is not accessible, an approximation is required. In the RPA the dielectric function is approximated by

\begin{equation}
\epsilon(\boldsymbol{r},\boldsymbol{r}',\omega)=\delta(\boldsymbol{r}-\boldsymbol{r}')-\int d\boldsymbol{r}'' v(\boldsymbol{r},\boldsymbol{r}'')P(\boldsymbol{r}'',\boldsymbol{r}',\omega),
\label{rpadiel1}
\end{equation}
where the polarization function $P(\boldsymbol{r}'',\boldsymbol{r}',\omega)$ is given by
\begin{equation}
\begin{gathered}
P(\boldsymbol{r},\boldsymbol{r}',\omega)=\\
\sum_{\sigma} \sum_{\boldsymbol{k},m}^{occ} \sum_{\boldsymbol{k}',m'}^{unocc} \varphi_{\boldsymbol{k}m}^{\sigma}(\boldsymbol{r}) \varphi_{\boldsymbol{k}'m'}^{\sigma*}(\boldsymbol{r}) \varphi_{\boldsymbol{k}m}^{\sigma*}(\boldsymbol{r}') \varphi_{\boldsymbol{k}'m'}^{\sigma}(\boldsymbol{r}')  \\
\times\Bigg[ \frac{1}{\omega-\Delta_{\boldsymbol{k}m,\boldsymbol{k}'m'}^{\sigma}} - \frac{1}{\omega+\Delta_{\boldsymbol{k}m,\boldsymbol{k}'m'}^{\sigma}} \Bigg].
\end{gathered}
\label{rpapol1}
\end{equation}
\newline
Here $\Delta_{\boldsymbol{k}m,\boldsymbol{k}'m'}^{\sigma}=\epsilon_{\boldsymbol{k}'m'}^{\sigma}-\epsilon_{\boldsymbol{k}m}^{\sigma}-i\eta$ 
with $\epsilon_{\boldsymbol{k}m}^{\sigma}$ the single particle Kohn-Sham eigenvalues obtained from DFT and $\eta$ a positive infinitesimal. 
Further, the $\varphi_{\boldsymbol{k}m}^{\sigma}(\boldsymbol{r})$ are the single particle Kohn-Sham eigenstates with 
spin $\sigma$, wavenumber $\boldsymbol{k}$ and band index $m$. The tags $occ$ and $unocc$ above the summation symbol indicate that the 
summation is respectively over occupied and unoccupied states only.

Eqs.~(\ref{fullysw}), (\ref{rpadiel1}), and (\ref{rpapol1}) constitute what is called the RPA of the dynamically 
screened Coulomb interaction. In the \textit{constrained} RPA the effective Coulomb interaction between a specific 
type of electrons in the system is considered. For example, in this work the effective Coulomb interaction between 
the 3$d$ electrons of iron will be investigated. Two types of RPA calculations are performed leading to fully and partially screened (effective $U$ or Hubbard $U$) Coulomb interaction parameters. In the 
latter the screening due to the electrons under consideration is excluded, i.e. in our case the 3$d$ electrons of 
iron. Thus, such a cRPA calculation provides the effective interaction that the electrons in the 3$d$ Hubbard model would experience; in other words, it yields the corresponding Hubbard $U$ parameter. 
Obviously it also gives insight to the importance of these 3$d$ electrons in the screening process.

In order to exclude the screening due to certain electrons one separates the polarization function in 
Eq.~(\ref{rpapol1}) as follows,
\begin{equation}
P=P_{l}+P_{r}.
\label{rpapol2}
\end{equation}
\newline
Here in our case $P_{l}$ includes only transitions between the strongly correlated 3$d$ states of iron and $P_{r}$ 
is the remainder. Then, the frequency dependent effective Coulomb interaction is given schematically by the matrix 
equation 
\begin{equation}
U(\omega) = [1-vP_{r}(\omega)]^{-1}v,
\end{equation}
where $v$ is the bare Coulomb interaction.

The problem with the separation of Eq.~(\ref{rpapol2}) is that it is only well defined for disentangled states. 
For entangled states different methods have been developed.\cite{ferd1,miy1,ers1} In this work we use the method described 
in Ref.~\onlinecite{ers1}. Here we first define the probability to find a strongly correlated electron 
(3$d$ state of iron in our case) in eigenstate $\varphi_{\boldsymbol{k}m}^{\sigma}$ as,
\begin{equation}
c_{\boldsymbol{k}m}^{\sigma}=\sum_{i,n} \lvert T_{i,mn}^{\sigma \boldsymbol{k}} \rvert^{2},
\label{probaen1}
\end{equation}
\newline
Here the unitary matrices $T_{i,mn}^{\sigma \boldsymbol{k}}$ are determined from the concept of maximally localized 
Wannier functions,
\begin{equation}
w_{in}^{\sigma}(\boldsymbol{r})=\frac{1}{N} \sum_{\boldsymbol{k}}e^{-i\boldsymbol{k} \cdot \boldsymbol{R}_{i}} \sum_{m} T_{i,mn}^{\sigma \boldsymbol{k}} \varphi_{\boldsymbol{k}m}^{\sigma}(\boldsymbol{r}),
\label{maxlocw1}
\end{equation}
where $w_{in}^{\sigma}(\boldsymbol{r})$ is a maximally localized Wannier function located at site $i$, $N$ 
is the number of discrete $k$ points in the full Brillouin zone and $\boldsymbol{R}_{i}$ the position vector 
of atomic site $i$. The matrices $T_{i,mn}^{\sigma \boldsymbol{k}}$ are determined by minimizing the spread of 
the Wannier functions,
\begin{equation}
\Omega=\sum_{i,n,\sigma}(\langle w_{in}^{\sigma} \lvert r^{2} \rvert w_{in}^{\sigma} \rangle - \langle w_{in}^{\sigma} \lvert \boldsymbol{r} \rvert w_{in}^{\sigma} \rangle^2).
\label{spread}
\end{equation}
Here the sum runs over all Wannier functions. It can be shown that the maximally localized Wannier functions 
constitute an orthonormal basis and that they resemble atomic orbitals, i.e. they are centered at an atomic 
site and decay with increasing distance from the site. Further, there is an efficient algorithm to find the 
$T_{i,mn}^{\sigma \boldsymbol{k}}$ under the condition that the spread is minimized. From Eq.~\ref{maxlocw1} 
it is clear that a choice has to be made on which bands to include for the construction of the maximally 
localized Wannier states. In practice (for entangled states), we make sure that enough bands are selected such that all the 
strongly correlated electron character is contained. Then, in general the number of maximally localized 
Wannier functions obtained from this space is larger than the dimensions spanned by the strongly correlated 
electrons. Therefore, a selection has to be made. Since the strongly correlated electrons are more localized 
than the other electrons, the idea is that the subset consisting of the most maximally localized Wannier 
functions correspond to the strongly correlated electrons.

For entangled states the probability $c_{\boldsymbol{k}m}^{\sigma} < 1$ in Eq.~(\ref{probaen1}), while for 
disentangled states $c_{\boldsymbol{k}m}^{\sigma}=1$. Then, the probability of an electron to be in the 3$d$ correlated subspace before and after a transition $\varphi_{\boldsymbol{k}m}^{\sigma} \rightarrow \varphi_{\boldsymbol{k}'m'}^{\sigma}$ is given by

\begin{equation}
p_{\boldsymbol{k}m\rightarrow\boldsymbol{k}'m'}^{\sigma}=c_{\boldsymbol{k}m}^{\sigma} c_{\boldsymbol{k}'m'}^{\sigma}.
\label{proben2}
\end{equation}
\newline
Thus for disentangled states $p_{\boldsymbol{k}m\rightarrow\boldsymbol{k}'m'}^{\sigma}=1$ and for entangled states $p_{\boldsymbol{k}m\rightarrow\boldsymbol{k}'m'}^{\sigma} < 1$. The polarization function $P_{l}$ now becomes    

\begin{equation}
\begin{gathered}
P_{l}(\boldsymbol{r},\boldsymbol{r}',\omega)=\\
\sum_{\sigma} \sum_{\boldsymbol{k},m}^{occ} \sum_{\boldsymbol{k}',m'}^{unocc}(p_{\boldsymbol{k}m\rightarrow\boldsymbol{k}'m'}^{\sigma})^{2} \varphi_{\boldsymbol{k}m}^{\sigma}(\boldsymbol{r}) \varphi_{\boldsymbol{k}'m'}^{\sigma*}(\boldsymbol{r}) \varphi_{\boldsymbol{k}m}^{\sigma*}(\boldsymbol{r}') \varphi_{\boldsymbol{k}'m'}^{\sigma}(\boldsymbol{r}')  \\
\times\Bigg[ \frac{1}{\omega-\Delta_{\boldsymbol{k}m,\boldsymbol{k}'m'}^{\sigma}} - \frac{1}{\omega+\Delta_{\boldsymbol{k}m,\boldsymbol{k}'m'}^{\sigma}} \Bigg].
\end{gathered}
\label{rpapolen2}
\end{equation}
\newline
By calculating the total polarization from Eq.~(\ref{rpapol1}) and $P_{l}$ from Eq.~(\ref{rpapolen2}), $P_{r}$ can be 
obtained from Eq.~(\ref{rpapol2}). For completeness, the effective Coulomb matrix within the selected subspace is computed by
\begin{equation}
\begin{gathered}
U_{in_{1},jn_{3},in_{2},jn_{4}}^{\sigma_{1},\sigma_{2}}(\omega)= \\
\int \int d\boldsymbol{r}d\boldsymbol{r}' w_{in_{1}}^{\sigma_{1}*}(\boldsymbol{r}) w_{jn_{3}}^{\sigma_{2}*}(\boldsymbol{r}') U(\boldsymbol{r},\boldsymbol{r}',\omega) w_{jn_{4}}^{\sigma_{2}}(\boldsymbol{r}') w_{in_{2}}^{\sigma_{1}}(\boldsymbol{r}).
\end{gathered}
\label{hubudef31}
\end{equation}
\newline
In this work we only consider the static limit ($\omega=0$). Furthermore, we use Slater parametrization,
\begin{equation}
\begin{gathered}
U_{i}=\frac{1}{(2l+1)^{2}}\sum_{m,m'} U_{im,im',im,im'}^{\sigma_{1},\sigma_{2}}(\omega=0) \quad \text{and} \\
V_{ij}=\frac{1}{(2l+1)^{2}}\sum_{m,m'} U_{im,jm',im,jm'}^{\sigma_{1},\sigma_{2}}(\omega=0).
\end{gathered}
\label{fhuburpa}
\end{equation}
\newline
Here $U_{i}$ is the effective onsite Coulomb interaction at site $i$ and $V_{ij}$ the effective intersite Coulomb 
interaction between sites $i$ and $j$. Note that although the matrix elements of the Coulomb potential are formally 
spin-dependent due to the spin dependence of the Wannier functions, we find that this dependence is negligible in practice.

\section{Computational details}
\label{section-3}

The DFT calculations are performed with the FLEUR code, which is based on a full-potential linearized 
augmented plane wave (FLAPW) implementation.\cite{fleur} All calculations are performed with an 
exchange-correlation functional in the generalized gradient approximation (GGA) as formulated by 
Perdew, Burke and Ernzerhof (PBE).\cite{gga} Further, all calculations are without spin orbit coupling.

Since it is a {\bf k}-space code, a supercell approach was employed for the cluster calculations, with 
a large empty space between clusters that were repeated in a periodic lattice. In our calculations a 
large unit cell of at least 12~\AA~dimensions is used in order to prevent the interaction between clusters 
of different unit cells. Further, for the cluster calculations the cutoff for the plane waves is  
3.6~Bohr$^{-1}$, $l_{cut}=8$ and the $\Gamma$ point is the only {\bf k}-point considered. The ground 
state geometric and magnetic structure of the Fe$_{2}$O$_{3}$,  Fe$_{3}$O$_{4}$ and Fe$_{4}$O$_{6}$ clusters 
is obtained from Refs.~\onlinecite{std8,std9} (see also Fig.~\ref{figclus}). More precisely, the geometries 
are optimized structures obtained from hybrid (B3LYP) functional calculations.\cite{hybrid} The Fe$_{2}$O$_{3}$ 
and Fe$_{4}$O$_{6}$ clusters are antiferromagnetic, while Fe$_{3}$O$_{4}$ is ferromagnetic.

For magnetite the geometric and magnetic structure is obtained from Refs.~\onlinecite{mag1,mag2}. 
Here the structure of magnetite is monoclinic with 56 atoms in the unit cell. The chemical formula 
is Fe$^{3+}_{A}$[Fe$^{2+}$,Fe$^{3+}]_{B}$O$_{4}$ with $A$ referring to tetrahedral sites occupied by 
Fe$^{3+}$ and $B$ to octahedral sites containing both Fe$^{2+}$ and Fe$^{3+}$. The magnetic moments 
of the $B$ sites are antiparallel to those of the $A$ sites. For the {\bf k}-mesh a grid of 6x6x2 equidistant {\bf k}-points is used. The cutoff for the plane waves is  4.0~Bohr$^{-1}$ and $l_{cut}=8$. 

The DFT calculations are used as an input for the SPEX code to perform RPA and cRPA calculations 
for the screened Coulomb interaction.\cite{spex} The SPEX code uses the Wannier90 library to construct 
the maximally localized Wannier functions.\cite{wan90,wan902} For this construction we used six 
states per iron atom, i.e. five 3$d$ states and one 4$s$ state.

\section{Results and Discussion}
\label{section-4}

\begin{figure}[!ht]
\begin{center}
\includegraphics[scale=0.45]{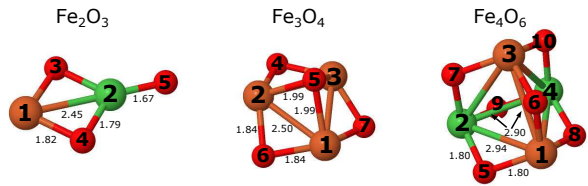}
\end{center}
\vspace{-1 cm}
\caption{The geometry of the  Fe$_{2}$O$_{3}$, Fe$_{3}$O$_{4}$ and Fe$_{4}$O$_{6}$ clusters. 
Here the red spheres correspond to the oxygen atoms, while the brown and green spheres correspond 
to iron atoms with antiparallel local magnetic moments. The distances between some atoms are 
provided in \AA. } 
\label{figclus}
\end{figure}

In Fig.\,\ref{figclus} the geometry and magnetic structure of the Fe$_{2}$O$_{3}$, Fe$_{3}$O$_{4}$ 
and Fe$_{4}$O$_{6}$ clusters is depicted. The red spheres correspond to the oxygen atoms, while 
the brown and green spheres correspond to iron atoms with antiparallel local magnetic moments. 
The distances between some of the atoms are given in \AA. Further, Fe$_{2}$O$_{3}$ and Fe$_{4}$O$_{6}$ 
are antiferromagnetic, while Fe$_{3}$O$_{4}$ is ferromagnetic. From Fe$_{4}$O$_{6}$ it can be observed 
that the direction of the local magnetic moment has a small influence on the bonding, i.e. the distance 
between two iron atoms with parallel and antiparallel moments is 2.90 and 2.94~\AA.

\begin{table}[!hbt]
\begin{center}
\caption{The bare and fully screened (RPA) average Coulomb interaction 
parameters for the Fe-3\emph{d} and O-2\emph{p} orbitals of the Fe$_{2}$O$_{3}$ cluster obtained from \emph{ab-initio} 
calculations. Here $U_{1}$ corresponds to the onsite Coulomb interaction of atom 1 and $V_{1,2}$ 
to the intersite Coulomb interaction between atoms 1 and 2 (see Fig.~\ref{figclus}). The second column indicates between what type of atoms this referes and the third column correponds to the distance in \AA{ }between them. Note that due to symmetry oxygen atoms 3 and 4 are equivalent. }
\begin{ruledtabular}
\begin{tabular}{lcccc}
U/V & Atom & Distance (\AA) & Bare (eV) & RPA (eV)   \\ \hline
$U_{1}$ & Fe & 0 & 21.7 & 7.7  \\ 
$U_{2}$ & Fe & 0 & 22.2 & 7.8  \\
$U_{3}$ & O & 0 & 17.8 & 8.2  \\
$U_{5}$ & O & 0 & 17.7 & 7.9  \\
$V_{2,5}$ & Fe-O & 1.67 & 8.6 & 6.7  \\
$V_{2,3}$ & Fe-O & 1.79 & 8.0 & 6.5  \\
$V_{1,3}$ & Fe-O & 1.82 & 7.8 & 6.4  \\
$V_{1,2}$ & Fe-Fe & 2.45 & 5.9 & 6.5  \\
$V_{3,4}$ & O-O & 2.66 & 5.6 & 6.0 \\
$V_{3,5}$ & O-O & 3.16 & 4.8 & 6.0  \\
$V_{1,5}$ & Fe-O & 4.11 & 3.9 & 6.0  \\
\end{tabular}
\end{ruledtabular}
\label{tabfe2o3}
\end{center}
\end{table}


\begin{table}[!hbt]
\begin{center}
\caption{The bare and partially screened (cRPA) average Coulomb interaction 
parameters for the Fe-3\emph{d} orbitals of the Fe$_{2}$O$_{3}$ cluster obtained from \emph{ab-initio} 
calculations. Here $U_{1}$ corresponds to the onsite Coulomb interaction of atom 1 and $V_{1,2}$ 
to the intersite Coulomb interaction between atoms 1 and 2 (see Fig.~\ref{figclus} to which atoms 
these numbers refer).}
\begin{ruledtabular}
\begin{tabular}{lcc}
U/V &Bare (eV) & cRPA (eV)  \\ \hline
$U_{1}$ & 21.7 & 8.7 \\ 
$U_{2}$ & 22.2 & 8.9 \\
$V_{1,2}$ & 5.9 & 6.3 \\
\end{tabular}
\end{ruledtabular}
\label{tabfe2o3c}
\end{center}
\end{table}

In the following first the matrix elements of the fully screened Coulomb interaction (RPA) of the Fe$_{x}$O$_{y}$ clusters will be discussed and second
the partially screened Coulomb interaction (cRPA) is briefly addressed. The latter is important in dealing with correlation effects in clusters as well as it provides information on the contribution of the Fe 
3\emph{d} $\rightarrow$ 3\emph{d} channel to the screening. In Table~\ref{tabfe2o3} the bare and fully screened onsite and intersite average Coulomb interaction parameters for Fe-3\emph{d} and O-2\emph{p} orbitals are presented for the smallest Fe$_{2}$O$_{3}$ cluster. Since oxygen atoms 3 and 4 are equivalent due to symmetry, only symmetry unequivalent interactions are presented. As seen the onsite Coulomb interactions are very well screened. On the other hand the intersite Coulomb interaction is much less screened and is more or less constant as function of intersite distance. Interestingly, starting from an intersite distance of $2.45$~\AA{ }anti-screening is observed, i.e. the fully screened interaction is larger than the bare interaction. For example, for the intersite Coulomb interaction between the two iron atoms the anti-screening contribution is 0.6~eV and between iron atom 1 and oxygen atom 5 it is even 2.1~eV. 

Table~\ref{tabfe2o3c} contains the partially screened (without iron 3$d$ contribution) average Coulomb interaction 
parameters for the Fe-3\emph{d} orbitals. As mentioned from this Table the contribution of the Fe-3\emph{d} electrons, i.e.  Fe 
3\emph{d} $\rightarrow$ 3\emph{d} channel, to the screening can be investigated. For the onsite Coulomb interaction this contribution is very small, 1~eV, compared to that of the Fe(3\emph{d}) $\rightarrow $ O(2\emph{p}) screening channel of about 13~eV. On the hand for the intersite iron-iron Coulomb interaction the Fe 
3\emph{d} $\rightarrow$ 3\emph{d} channel contributes significantly, 0.2~eV, to the total anti-screening effect of 0.6~eV.

\begin{table}[!hbt]
\begin{center}
\caption{The same as in Table\,\ref{tabfe2o3} for the Fe$_{3}$O$_{4}$ cluster. Note that due to symmetry the iron atoms are equivalent, while for oxygen atoms 4, 6 and 7 are equivalent. }
\begin{ruledtabular}
\begin{tabular}{lcccc}
U/V & Atom & Distance (\AA) & Bare (eV) & RPA (eV)   \\ \hline
$U_{1}$ & Fe & 0 & 22.2 & 7.4  \\ 
$U_{4}$ & O & 0 & 17.8 & 7.8  \\
$U_{5}$ & O & 0 & 17.9 & 8.1  \\
$V_{1,6}$ & Fe-O & 1.84 & 7.8 & 5.8  \\
$V_{1,5}$ & Fe-O & 1.99 & 7.2 & 5.7  \\
$V_{1,2}$ & Fe-Fe & 2.50 & 5.8 & 5.8  \\
$V_{4,5}$ & O-O & 2.73 & 5.4 & 5.3  \\
$V_{4,6}$ & O-O & 3.40 & 4.5 & 5.1 \\
$V_{1,4}$ & Fe-O & 3.45 & 4.4 & 5.3  \\
\end{tabular}
\end{ruledtabular}
\label{tabfe3o4}
\end{center}
\end{table}

\begin{table}[!t]
\begin{center}
\caption{The same as in Table\,\ref{tabfe2o3c} for the Fe$_{3}$O$_{4}$ cluster. 
%
%
Note that due to symmetry only one onsite and intersite interaction is presented.}
\begin{ruledtabular}
\begin{tabular}{lcc}
U/V &Bare (eV)& cRPA (eV)  \\ \hline
$U_{1}$  & 22.2 & 8.6  \\
$V_{1,2}$ & 5.8  & 5.5  \\
\end{tabular}
\label{tabfe3o4c}
\end{ruledtabular}
\end{center}
\end{table}

A similar behavior can be observed for Fe$_{3}$O$_{4}$ and Fe$_{4}$O$_{6}$. Their results for the 
bare and fully screened average Coulomb interaction parameters for the Fe-3\emph{d} and O-2\emph{p} 
orbitals are shown in Tables~\ref{tabfe3o4} and \ref{tabfe4o6}. The partially screened average Coulomb interaction parameters for the Fe-3\emph{d} orbitals are presented  in Tables~\ref{tabfe3o4c} and \ref{tabfe4o6c}. In Fe$_{3}$O$_{4}$ iron atoms 1, 2, 3 and oxygen atoms 4, 6, 7 are equivalent due to symmetry. Therefore, only symmetry unequivalent onsite and intersite interactions are shown in Tables~\ref{tabfe3o4} and \ref{tabfe3o4c}. From Table~\ref{tabfe3o4} it can again be observed that the onsite Coulomb interaction is very well screened, while the intersite Coulomb interaction is almost constant throughout the cluster. Furthermore, anti-screening is again present although it starts to occur at a larger intersite distance. 

From a comparison of Tables~\ref{tabfe3o4} and \ref{tabfe3o4c} it is also observed for this cluster that the screening contribution of the Fe 
3\emph{d} $\rightarrow$ 3\emph{d} channel to the onsite Coulomb interaction is small, 1.2~eV, compared to that of the \mbox{Fe(3\emph{d}) $\rightarrow $ O(2\emph{p})} screening channel, 13.6~eV. Further, it appears that there is an anti-screening contribution of 0.3~eV of the \mbox{Fe 3\emph{d} $\rightarrow$ 3\emph{d}} channel to the iron-iron intersite interaction.

For the Fe$_{4}$O$_{6}$ cluster the iron atoms 1, 3 and 2, 4 are related by symmetry, while for oxygen 5, 7, 8, 10 and 6, 9 are related. Also for this cluster it can be observed from Table~\ref{tabfe4o6} that the onsite Coulomb interaction is very well screened, while the intersite Coulomb interaction is almost constant in the cluster. The anti-screening in this cluster is a bit different. It occurs between two irons at intersite distances of 2.90 and 2.94~\AA{ }, while it is absent between two oxygen atoms until an intersite distance of 4.13~\AA. Between iron and oxygen anti-screening starts at an intersite distance of 3.43~\AA.   
 
As for the other clusters, the screening contribution of the Fe 3\emph{d} $\rightarrow$ 3\emph{d} channel to the onsite Coulomb interaction is small, 1.0~eV (see Table~\ref{tabfe4o6c}), compared to that of the Fe(3\emph{d}) $\rightarrow $ O(2\emph{p}) screening channel, 15.4~eV. However, in contrast to the other clusters it appears from Table~\ref{tabfe4o6c} that there is no anti-screening contribution from the Fe 3\emph{d} $\rightarrow$ 3\emph{d} channel.

\begin{table}[!hbt]
\begin{center}
\caption{The same as in Table~\ref{tabfe2o3} for the Fe$_{4}$O$_{6}$ cluster. Note that due to symmetry the iron atoms 1, 3 and 2, 4 are equivalent, while for oxygen atoms 5, 7, 8, 10 and 6, 9 are equivalent. }
\begin{ruledtabular}
\begin{tabular}{lcccc}
U/V & Atom & Distance (\AA) & Bare (eV) & RPA (eV)   \\ \hline
$U_{1}$ & Fe & 0 & 22.3 & 5.9  \\ 
$U_{3}$ & Fe & 0 & 22.3 & 5.9  \\
$U_{5}$ & O & 0 & 18.1 & 6.8  \\
$U_{6}$ & O & 0 & 18.0 & 7.0  \\
$V_{1,5}$ & Fe-O & 1.80 & 8.0 & 5.1  \\
$V_{1,6}$ & Fe-O & 1.83 & 7.8 & 5.0  \\
$V_{1,3}$ & Fe-Fe & 2.90 & 5.1 & 5.3  \\
$V_{5,7}$ & O-O & 2.92 & 5.1 & 4.8 \\
$V_{1,2}$ & Fe-Fe & 2.94 & 5.0 & 5.2  \\
$V_{5,6}$ & O-O & 3.00 & 5.0 & 4.8  \\
$V_{5,9}$ & O-O & 3.00 & 5.0 & 4.8  \\
$V_{1,7}$ & Fe-O & 3.43 & 4.4 & 5.0  \\
$V_{1,9}$ & Fe-O & 3.54 & 4.3 & 4.9  \\
$V_{5,10}$ & O-O & 4.13 & 3.8 & 4.7  \\
$V_{6,9}$ & O-O & 4.34 & 3.7 & 4.7  \\
\end{tabular}
\end{ruledtabular}
\label{tabfe4o6}
\end{center}
\end{table}

\begin{table}[!hbt]
\begin{center}
\caption{The same as in Table~\ref{tabfe2o3c} for the Fe$_{4}$O$_{6}$ cluster. Due to symmetry iron atoms 1, 3 and 2, 4 are equivalent.
%
%
}
\begin{ruledtabular}
\begin{tabular}{lcc}
U/V &Bare (eV) & cRPA (eV)   \\ \hline 
$U_{1}$  & 22.3 & 6.9  \\
$U_{3}$  & 22.3 & 6.9  \\
$V_{1,2}$ & 5.0 & 5.2 \\
$V_{1,3}$ & 5.1 & 5.4  \\
\end{tabular}
\label{tabfe4o6c}
\end{ruledtabular}
\end{center}
\end{table}

Previous studies have shown that anti-screening strongly manifest itself in low-dimensional semiconductors 
and insulators.\cite{mot1,CNT} Using a point-dipole interaction model van den Brink and Sawatsky calculated 
the screened intersite Coulomb interaction for finite size systems like molecules (benzene, naphtaline, 
C$_{60}$, etc.) and one-dimensional atomic chains.\cite{mot1} The authors found that, in contrast to three-dimensional 
bulk semiconductors, in low-dimensional systems the local field effects play a very important role in screening 
of the Coulomb interaction. It turns out that the Coulomb interaction is strongly $r$-dependent, i.e., at short 
distances it is strongly screened, at intermediate distances it is anti-screened, and at large distances it 
is unscreened. The occurrence of anti-screening in low-dimensional systems was attributed to the sign change 
of the induced polarization around the test charge with distance. In three-dimensional insulators and semiconductors 
the induced polarization is negative over all space, while in low-dimensional systems it can change sign 
with distance resulting in an anti-screening. The critical distance $r_c$, where the transition from screening 
to anti-screening takes place, depends very much on the dimensionality and polarization of the system. For instance, 
in zero-dimensional molecules (benzene, naphthalene) $r_c$ is rather small, 3-4~\AA \cite{mot1}, while in quasi-one 
dimensional single-wall carbon nanotubes it is around 20~\AA.\cite{CNT} For the Fe$_{x}$O$_{y}$ clusters
considered in the present work the critical distance $r_c$ can be even shorter than the zero dimensional 
systems studied in literature. For Fe$_{2}$O$_{3}$ and Fe$_{3}$O$_{4}$ the critical distance is respectively about 2.45 and 3.40~\AA. The critical distance for Fe$_{4}$O$_{6}$ is a bit unambiguous. Namely anti-screening is first observed at an intersite distance of 2.90~\AA, then both screening and anti-screening are observed until an intersite distance of 3.43~\AA. Note that even in three-dimensional bulk materials the non-local anti-screening 
takes place within the sub-space of the correlated electrons as recently shown by Nomura \textit{et al.}, for 
the case of SrVO$_3$.~\cite{SrVO3}

\begin{table}[!hbt]
\begin{center}
\caption{The bare, partially screened (cRPA) and fully screened (RPA) average Coulomb interaction parameters  for 
the Fe-3\emph{d} orbitals of magnetite obtained from \emph{ab-initio} calculations. Here the first column 
shows on or between which sublattice the interaction is considered and the second column contains the distance 
between these sublattices (a zero referring to an onsite interaction). For the sublattices the same nomenclature 
is adopted as in Ref.~\onlinecite{mag1}.}
\begin{ruledtabular}
\begin{tabular}{lcccc}
& $r$(\AA) &Bare (eV) & cRPA (eV) & RPA (eV)  \\ \hline
A1      & 0     & 22.9 &  4.3  & 1.53   \\
A2      & 0     & 22.9 &  4.3  & 1.51  \\
B1a     & 0     & 22.9 &  4.8  & 0.75  \\
B1b     & 0     & 22.9 &  4.8  & 0.77  \\
B2a     & 0     & 22.9 &  4.7  & 0.82  \\
B3      & 0     & 22.9 &  4.6  & 0.81  \\
A1-A2   & 6.93 & 2.4  &  0.01 & 0.01  \\
B1a-B2a & 5.10  & 3.0  &  0.09 & 0.02  \\
B1a-B1b & 2.97  & 4.9  &  0.35 & 0.02 \\
B1b-B3  & 2.86  & 5.1  &  0.37 & 0.04  \\
\end{tabular}
\label{tabmagn}
\end{ruledtabular}
\end{center}
\end{table}

It is interesting to compare these cluster results with their bulk counterpart magnetite. In Table~\ref{tabmagn} 
the calculated results are shown for magnetite. Here the first column shows on or between which sublattices 
the interaction is considered and the second column contains the distance between these sublattices (a zero 
indicates an onsite interaction). For the sublattices the same nomenclature is adopted as in Ref.~\onlinecite{mag1}. 
From Table~\ref{tabmagn} it can be observed that the intersite Coulomb interaction is almost completely screened, 
which is in strong contrast with the zero-dimensional cluster results. The onsite Coulomb interaction 
is also more screened than for the clusters. Furthermore, the cRPA calculations reveal that the effect of the 
screening due to the iron 3$d$ states is quite substantial in magnetite. 

Finally we would like to comment on the strength of the electronic correlations in three-dimensional magnetite Fe$_3$O$_4$ 
and zero-dimensional Fe$_{x}$O$_{y}$ clusters. The short range nature of the Coulomb interaction with large gradient
in magnetite makes it a correlated material and thus electronic structure methods which go beyond the standard 
DFT are necessary for an accurate description of the electronic structure of magnetite. For instance, the 
experimentally observed charge order in magnetite cannot be captured in DFT. From Ref.~\onlinecite{mag1} it is known that an additional treatment of the onsite correlations between the Fe 3$d$ electrons is needed. It was found that the DFT+U approach, a static mean-field treatment of onsite correlations, gives a charge ordering in agreement with experiment.  
On the other hand, due to the almost constant Coulomb interaction in zero-dimensional 
Fe$_{x}$O$_{y}$ clusters, DFT calculations employing standard functionals can be expected to capture the essential physics. For example, from a comparison of the experimental vibrational spectrum with the theoretical spectra of different isomers and magnetic structures, the geometric and magnetic structure are obtained in good agreement with the experiment.\cite{std7,std9}

\section{Conclusion}
\label{section-5}

We have performed RPA and cRPA calculations for the effective Coulomb interaction in the Fe$_{2}$O$_{3}$, Fe$_{3}$O$_{4}$ and Fe$_{4}$O$_{6}$ clusters and their bulk counterpart magnetite. It has been demonstrated that both in the clusters and bulk the onsite Coulomb interaction is very well screened. Here the main screening contribution stems from the Fe(3\emph{d}) $\rightarrow $ O(2\emph{p}) channel. On the other hand the intersite Coulomb interaction is barely screened or even anti-screened in the clusters, while in the bulk it is almost completely screened. In Fe$_{2}$O$_{3}$ and Fe$_{3}$O$_{4}$ the anti-screening starts at a certain intersite distance, 2.45 and 3.40~\AA{ }respectively. For Fe$_{4}$O$_{6}$ the anti-screening nature is a bit more complex. It first occurs at a distance of 2.45~\AA{ }, then both screening and anti-screening can be observed until a distance of 3.43~\AA{ }from which on it is of anti-screening nature only. The important consequence is that in the clusters the Coulomb interaction is almost constant, while in the bulk it has a large gradient. Therefore, a proper treatment of correlations are expected to be more important for the bulk than the clusters. 

\subsection*{Acknowledgements}
The Nederlandse Organisatie voor Wetenschappelijk Onderzoek (NWO) and SURFsara 
are acknowledged for the usage of the LISA supercomputer and their support.L.P. and
M.I.K. acknowledges a support by European ResearchCouncil (ERC) Grant No. 338957.



\end{document}